\documentclass[12pt, draftcls, onecolumn, letterpaper]{IEEEtran}

\ifCLASSINFOpdf

\else

\fi

\ifCLASSOPTIONcompsoc
    \usepackage[caption=false, font=normalsize, labelfont=sf, textfont=sf]{subfig}
\else
    \usepackage[caption=false, font=normalsize]{subfig}
\fi
\usepackage{lipsum}%
\usepackage[dvipsnames]{xcolor}
\usepackage{algorithm,algorithmic}
\usepackage{balance}
\usepackage{multicol}   
\usepackage{cite}
\usepackage{gensymb}
\usepackage{multirow}
\usepackage{graphics}
\usepackage{epsfig}
\usepackage{graphicx}
\usepackage{epstopdf}
\usepackage{textcomp}
\usepackage{amsmath}
\usepackage{mathtools}
\usepackage{filecontents}
\usepackage{lipsum,color}
\usepackage{amssymb}
\usepackage{float}
\usepackage{colortbl} 
\usepackage{times} 
\usepackage{amsthm}  

\usepackage{amsfonts}

\theoremstyle{break}

\begin{document}
\title{\LARGE Pointing Error Modeling of mmWave to THz High-Directional Antenna Arrays}
\author{ Mohammad~Taghi~Dabiri,~Mazen~Hasna,~{\it Senior Member,~IEEE}
	\thanks{This publication was made possible by grant number NPRP13S-0130-200200 from the Qatar National Research Fund, QNRF. The statements made herein are solely the responsibility of the authors.}
	\thanks{The authors are with the Department of Electrical Engineering, Qatar University, Doha, Qatar.  (E-mail: m.dabiri@qu.edu.qa; hasna@qu.edu.qa).}
}


\maketitle
\begin{abstract}
This paper focuses on providing an analytical framework for the quantification and evaluation of the pointing error at high-frequency millimeter wave (mmWave) and terahertz (THz) communication links. For this aim, we first characterize the channel of a point-to-point communication link between unstable transmitter (Tx) and receiver (Rx) and then, we derive the probability density function (PDF) and cumulative distribution functions (CDF) of the pointing error in the presence of unstable Tx and Rx as a function of the antennas' pattern. Specifically, for the standard array antenna, a closed-form expression is provided for the PDF of the pointing error, which is a function of the number of antenna elements. Moreover, a more tractable approximate model is provided for the CDF and PDF of pointing error. In addition, using $\alpha-\mu$ distribution, which is a common model for small-scale fading of THz links, the end-to-end PDF of the considered channel is derived and used to calculate the outage probability of the considered system. Finally, by employing Monte-Carlo simulations, the accuracy of the analytical expressions is verified and the performance of the system is studied.
\end{abstract}
\begin{IEEEkeywords}
Antenna arrays, misalignment, backhaul links, pointing errors, mmWave, THz systems.
\end{IEEEkeywords}
\IEEEpeerreviewmaketitle

\section{Introduction}

\begin{figure*}
	\centering
	\subfloat[] {\includegraphics[width=3.2 in]{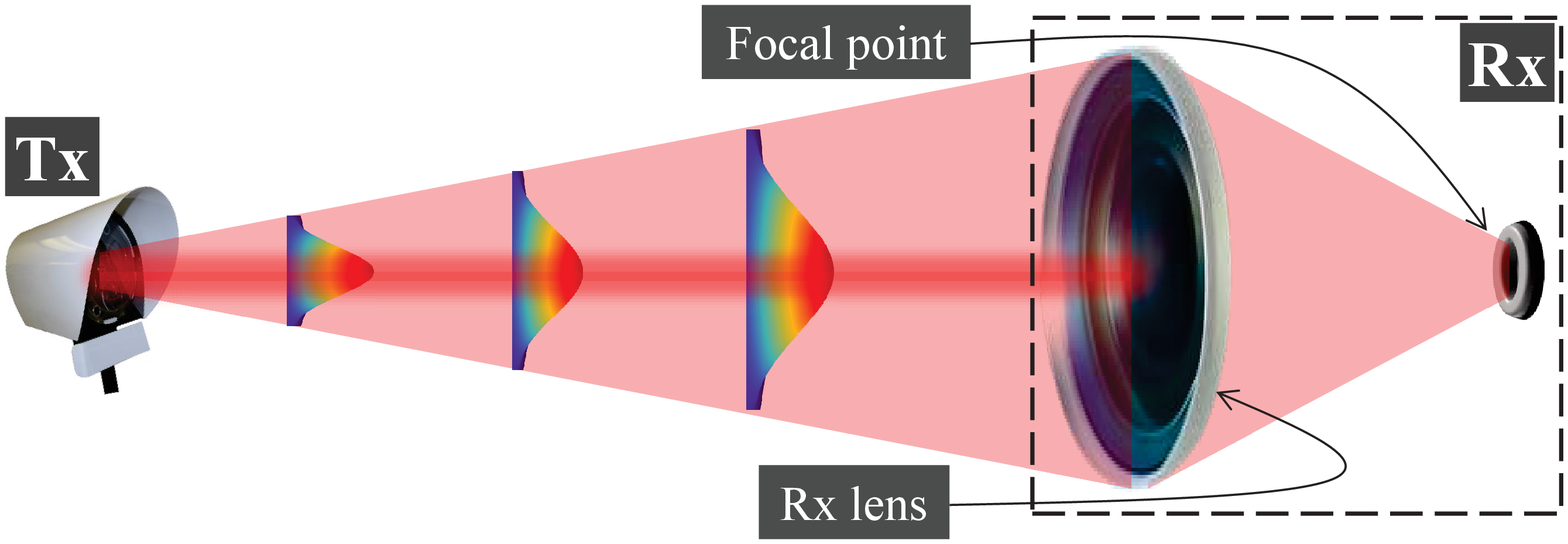}
		\label{n1}
	}
	\hfill
	\subfloat[] {\includegraphics[width=4.2 in]{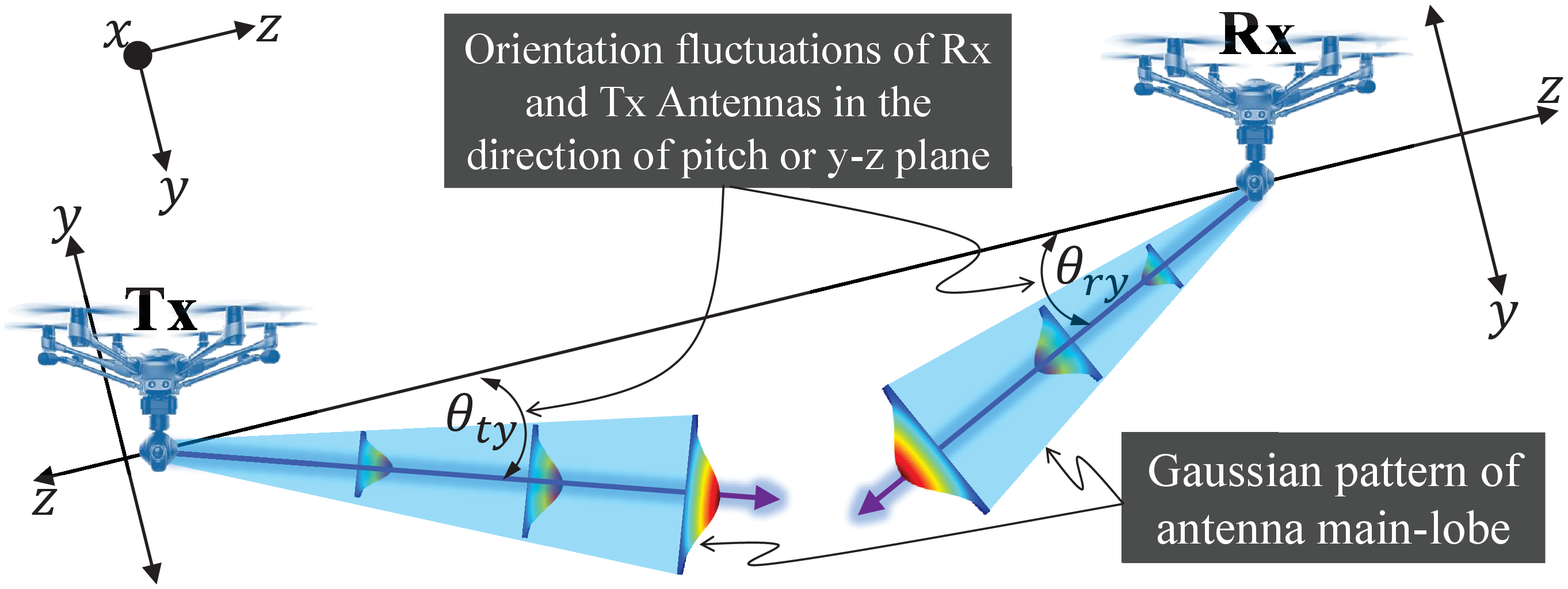}
		\label{n2}
	}
	\hfill
	\subfloat[] {\includegraphics[width=1.5 in]{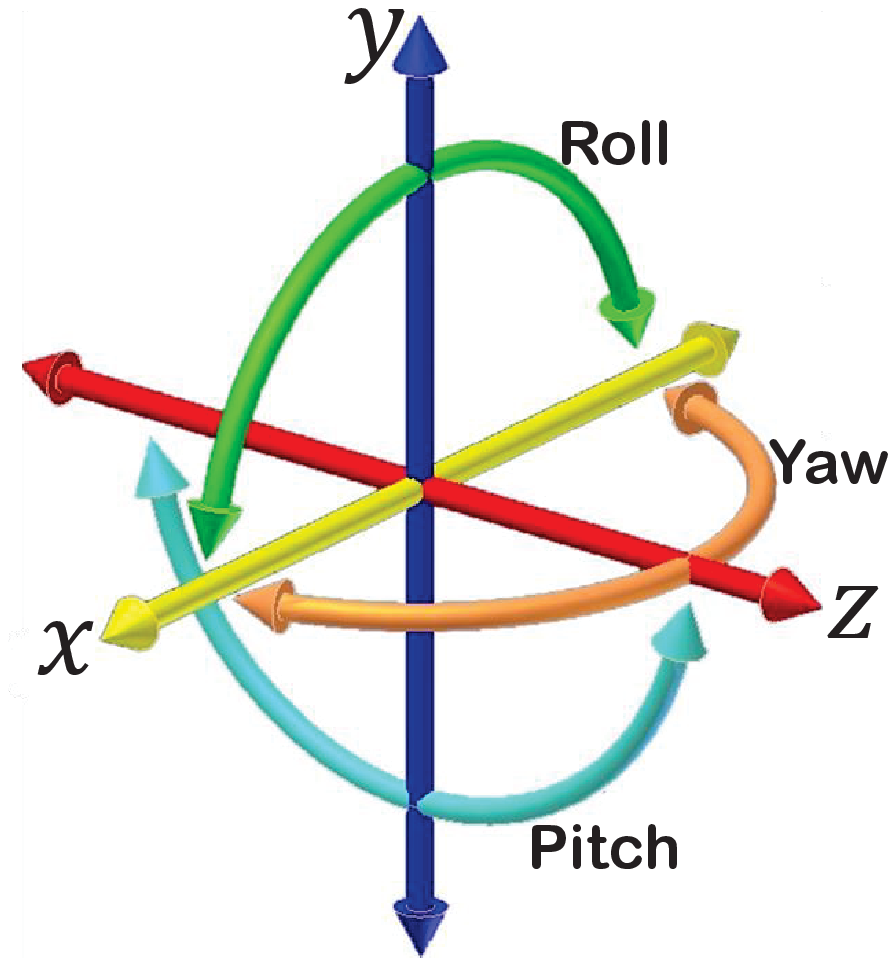}
		\label{n3}
	}
	\caption{Graphical illustration of  (a)  an optical Rx system that the aperture lens collects the received Gaussian laser beam and focuses it on the photodetector; 
		(b) pointing errors of two directional antennas which are mounted on unstable Tx and Rx;
		(c) the relationship between the axes of the Cartesian coordinate system and the roll, yaw, and pitch directions in the considered system model. }
	\label{n4}
\end{figure*}

\IEEEPARstart{T}{oday}, high-frequency millimeter wave (mmWave) and terahertz (THz) communication links are widely considered as the next frontiers for future wireless systems. 
While mmWaves are already used in 5G, the research community focus is gradually shifting to taking advantage of the enormous bandwidth of higher THz frequencies.
\textcolor{black}{Communicating at such high frequencies comes with a drawback of high path-loss. However, the small wavelength enables the realization of a compact form of high directional antenna arrays. In terrestrial mmWave/THz links, communication with narrow beams might easily suffer from blockage, and hence, employing aerial nodes offer a higher probability of line-of-sight (LoS) between communication nodes. One of the attractive scenarios is to use the large available bandwidth at mmWave/THz frequencies in order to provide extra data rate for point-to-point aerospace communications. In practical situations, an error in mechanical control systems, mechanical noise, position estimation errors, air pressure, and wind speed can affect the angular and position stability of aerial nodes, especially for small unmanned aerial vehicles (drones). Potentially, high-directional mmWave/THz antennas suffer from transceivers antenna misalignment, well-known as pointing error, which affects the strength of the received signal and deteriorates the link performance. Therefore, along with other research topics in mmWave/THz bands, studying the effect of pointing errors on system performance is a necessity for establishing a high reliable communication link, which has recently been the subject of several studies \cite{badarneh2022performance,9492775,du2022performance,boulogeorgos2019analytical,9217121,9614341,boulogeorgos2022joint}. In summary, in \cite{badarneh2022performance,9492775,du2022performance,boulogeorgos2019analytical,9217121,9614341,boulogeorgos2022joint}, the well-known pointing error model provided in \cite{farid2007outage} is used, which, although suitable for optical communication systems and a special case of mmWave/THz systems, as we will show in this work, cannot be directly used for the typical mmWave/THz communication systems. n particular, in an optical system, only the effect of angular fluctuations of the transmitter (Tx) is considered and the angular fluctuations of the receiver (Rx) do not affect the pointing error model provided in \cite{farid2007outage}. In fact, for a lens-based Rx system, the angular fluctuations of the Rx cause fluctuations of the focused beam at the focal plane of the converging lens \cite{dabiri2018channel}, which is called the beam wandering at the focal plane and is different from the geometrical pointing error provided in \cite{farid2007outage}. Therefore, to quantify the benefits of using mmWave/THz frequencies for aerial/mobile communication systems, it is essential to have an accurate channel modeling that incorporates the effect of pointing error based on the severity of Tx and Rx vibrations, which is our main motivation for doing this research work.}

%
%
%

We first characterize the channel of a point-to-point communication link between unstable Tx and Rx by taking into account the effects of transceiver vibrations, three-dimensional (3D) real antenna pattern, path loss and small-scale fading. 
%
%
Also, for the standard array antenna pattern, an approximate Gaussian model is presented based on the number of antenna elements. By using this model, we derive the probability density function (PDF) and cumulative distribution functions (CDF) of the pointing error in the presence of unstable Tx and Rx as a function of the antenna pattern.
Moreover, a more tractable approximate model is provided for the CDF and PDF of pointing error. 
In addition, using $\alpha-\mu$ distribution, which is a commonly used model for small-scale fading of THz links, the end-to-end PDF of the considered channel is derived. Finally, by employing Monte-Carlo simulations, the accuracy of the analytical expressions is verified and the performance of the system is studied.


\section{The System Model}
%
Consider the Rx in Fig. \ref{n1} where a lens is used to focus the incoming received signal onto a small detector or feed antenna located at the focal point. 
In this case, if the center of the received Gaussian beam deviates by $d_v$ from the center of the Rx lens, the fraction of the collected power by the the circular lens with radius $a$ (called the pointing error) is calculated as \cite{farid2007outage}
\begin{align}
	\label{hp1}
	h_\text{p} = \int_{-a}^a \int_{-\sqrt{a^2-x^2}}^{\sqrt{a^2-x^2}} \frac{2}{\pi w_z^2}
	\exp\left(     -2 \frac{(x-d_v)^2+y^2}{w_z^2}  
	\right)   \text{d}x\text{d}y,
\end{align}
where $w_z$ is the beamwidth of the received Gaussian beam. Using \eqref{hp1} and after a series of manipulations, the distribution of $h_p$ is derived in \cite[Eq. (11)]{farid2007outage}.
As mentioned earlier, in the pointing error studies of THz communications, it is typical to refer to \cite[Eq. (11)]{farid2007outage} in order to model the pointing error. Although this model is suitable for free-space optical (FSO) systems and/or a specific class of THz communication links that use a circular lens in the Rx, in this work, we show that this model is not an accurate model for Lens-less THz communication systems.

Consider next Fig. \ref{n2}, where we consider a point-to-point THz communication system between two unstable nodes, and where Tx and Rx are equipped with high directional antennas. The channel model can be written as \cite{kokkoniemi2020impact,balanis2016antenna}
\begin{align}
	\label{he1}
	h           =   h_L h_a \sqrt{G_t(\theta_t,\phi_t) G_r(\theta_r,\phi_r)},
\end{align}
where $h_a$ is the small scale fading, $h_L=h_{Lf}h_{Lm}$ is the channel path loss, $h_{Lf}=\left(\frac{\lambda}{4\pi Z}\right)^2$ is the free-space path loss, $h_{Lm}= e^{-\frac{\mathcal{K}(f)}{2}Z}$ denotes the molecular absorption loss where $\mathcal{K}(f)$ is the frequency dependent absorption coefficient, and $Z$ is the link length.
The experimental results show that water vapor dominates the molecular absorption loss at high THz frequencies \cite{tripathi2021millimeter}.
In THz channels, the $\alpha-\mu$ distribution is one of the most widely used models employed for the distribution function of random variable (RV) $h_a$ which is given as \cite{papasotiriou2021experimentally}
\begin{align}
	\label{he3}
	f_{h_a}(h_a) = \frac{\alpha \mu^\mu}{\hat{h_a}^{\alpha \mu} \Gamma(\mu)} h_a^{\alpha \mu -1}  \exp\left( -\mu \frac{ h_a^\alpha}{\hat{h_a}^\alpha}  \right),
\end{align}
where $\alpha>0$ is a fading parameter, $\mu$ is the normalized variance of the fading channel envelope, and $\hat{h_a}$ is the $\alpha$-root mean value of the fading channel envelope.

In \eqref{he1}, the parameter $G_q(\theta_q,\phi_q)$ is the antenna radiation pattern in the directions of $\theta_q$ and $\phi_q$ where the subscript $q\in\{t,r\}$ determines the Tx and Rx nodes.  
Here, we consider a standard uniform $N\times N$ array antenna. 
%
%
By taking into account the effect of all elements, the array radiation gain in the direction of $\theta_q$ and $\phi_q$ will be 
$G_q(\theta_q, \phi_q) = G_0(N) G'_q(\theta_q, \phi_q)$, where
\cite{balanis2016antenna} 
\begin{align}
	\label{f_1}
	&G'_q(\theta_q, \phi_q) = \\
	&\left( \frac{\sin\left(\frac{N (k d_{x} \sin(\theta_q)\cos(\phi_q))}{2}\right)} 
	{N\sin\left(\frac{k d_{x} \sin(\theta_q)\cos(\phi_q)+\beta_{x}}{2}\right)}
	 \frac{\sin\left(\frac{N (k d_{y} \sin(\theta_q)\sin(\phi_q))}{2}\right)} 
	{N\sin\left(\frac{k d_{y} \sin(\theta_q)\sin(\phi_q)}{2}\right)}\right)^2, \nonumber
\end{align}
$d_{x}=d_y=\frac{\lambda}{2}$ are the spacing between the elements along the  $x$ and $y$ axes, respectively,
$k=\frac{2\pi}{\lambda}$ denotes the wave number, and $\lambda=\frac{c}{f_c}$ denotes the wavelength, where $f_c$ denotes the carrier frequency and $c$ is the speed of light. 
Also, in order to guarantee that the total radiated power of antennas with different $N$ are the same, the coefficient $G_0$ is defined as \cite{balanis2016antenna}
\begin{align}
	\label{cv}
	G_0(N)={4\pi}\left( {\int_0^{\pi}\int_0^{2\pi} G'_q(\theta_q,\phi_q) \sin(\theta_q) d\theta_q d\phi_q} \right)^{-1}.
\end{align}
Based on \eqref{f_1}, the maximum value of the antenna gain is equal to $G_0(N)$, which is obtained when $\theta_q=0$.

As shown in Fig. \ref{n2}, we assume that the Tx and Rx are located on axis $z$ and at a distance $Z$ from each other, and both the Tx and Rx try to place the main lobe of the antenna pattern on the $z$ axis. The use of THz high-gain antennas makes them more sensitive to antenna misalignment or pointing errors, especially for mobile or aerospace communications. 
The angular fluctuations of each mobile node are modeled in three directions; namely: yaw, roll, and pitch. Due to the symmetry in the main-lobe of the antenna pattern in the direction of roll (angle $\phi_q$), we can well neglect the effect of orientation deviations in the roll direction on the pointing error. Hence,
the orientation fluctuations in the yaw and pitch directions cause a pointing error. 
As shown in Fig. \ref{n3}, orientation deviations in the directions of yaw and pitch are equivalent to the orientation deviations in $x-z$ and $y-z$ planes, respectively. Let $\theta_{tx}\sim\mathcal{N}(0,\sigma_\theta)$ and $\theta_{ty}\sim\mathcal{N}(0,\sigma_\theta)$ denote the orientation fluctuations of Tx in $x-z$ and $y-z$ planes, respectively, and $\theta_{rx}\sim\mathcal{N}(0,\sigma_\theta)$ and $\theta_{ry}\sim\mathcal{N}(0,\sigma_\theta)$ denote the orientation fluctuations of Rx in $x-z$ and $y-z$ planes, respectively. 
Therefore, in our model, the RVs $\theta_{q}$ and $\phi_q$ can be defined as  functions of RVs $\theta_{qx}$ and $\theta_{qy}$ as follows \cite{dabiri2020analytical}:
\begin{align}
	\label{f_2}
	\theta_q  &= \tan^{-1}\left(\sqrt{\tan^2(\theta_{qx})+\tan^2(\theta_{qy})}\right).
\end{align}

\section{Pointing Error Modeling}
\begin{figure}
	\centering
	\subfloat[] {\includegraphics[width=1.75 in,height= 1.1 in]{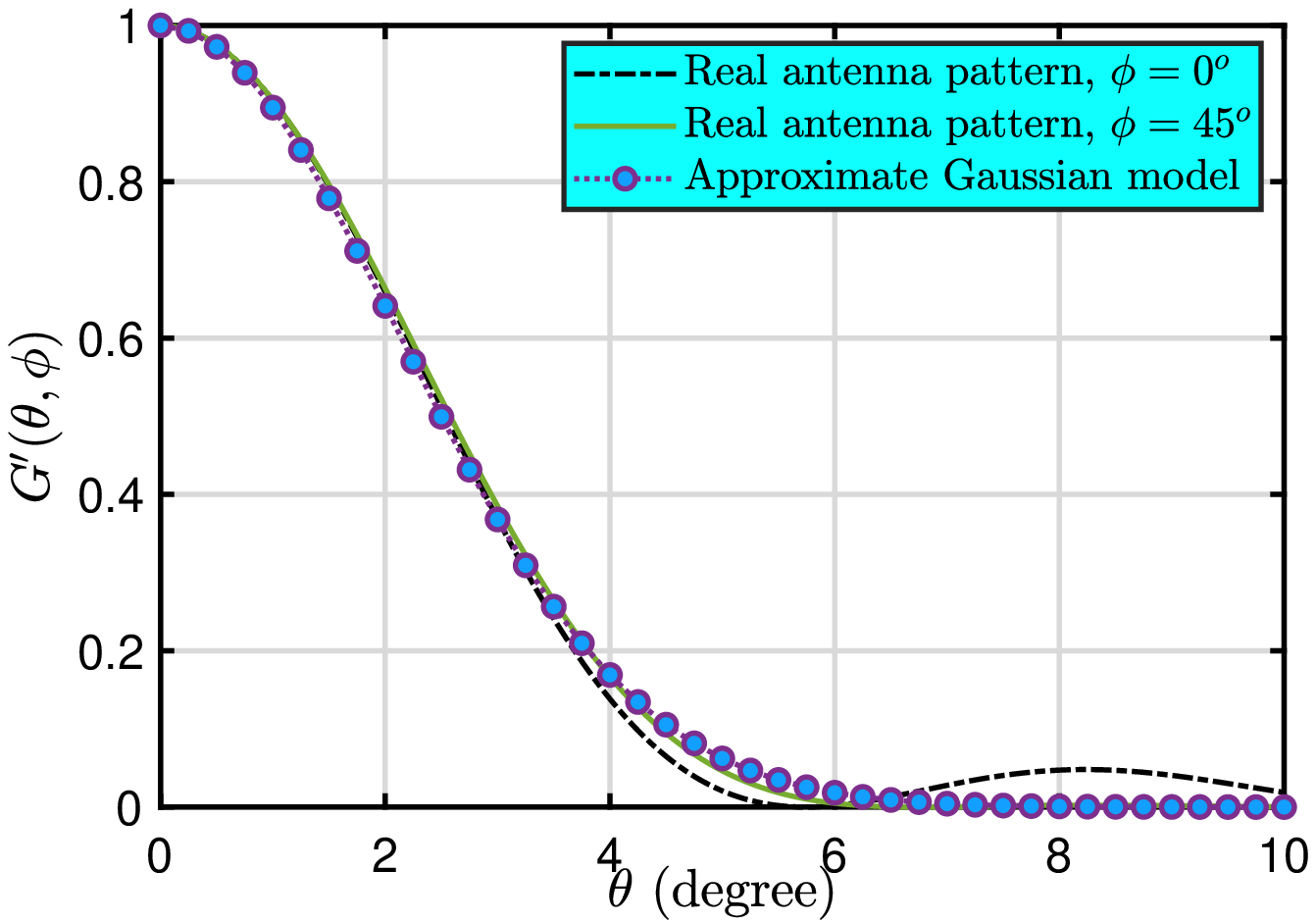}
		\label{m1}
	}
	\subfloat[] {\includegraphics[width=1.75 in,height= 1.1 in]{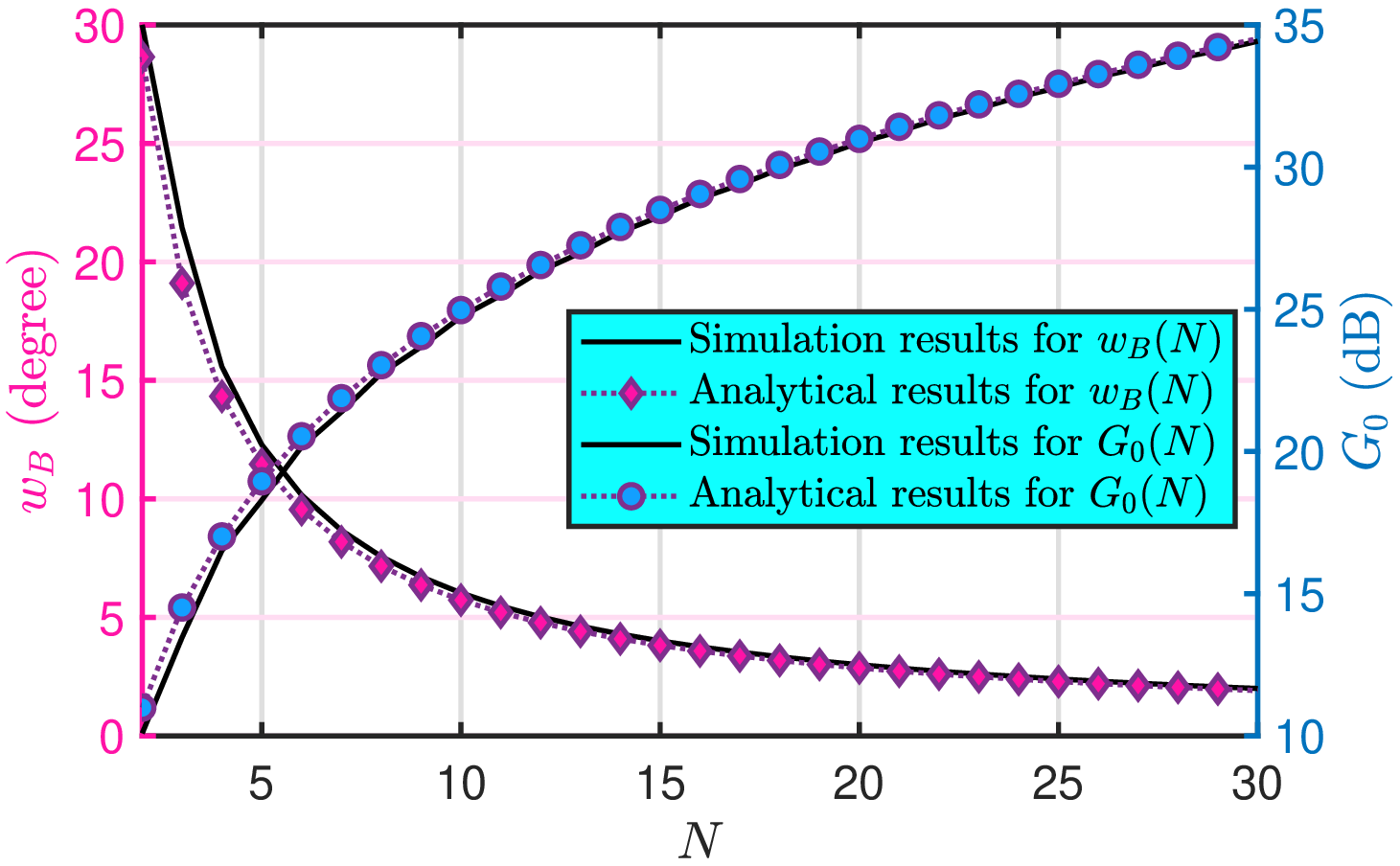}
		\label{m2}
	}
	\caption{(a) Comparison of Gaussian approximate pattern with actual antenna pattern of an array antenna for two different values of angle $\phi$.
		(b) Comparing the accuracy of approximate values for parameters $w_B$ and $G_0$ with the values obtained from numerical results.}
	\label{m3}
\end{figure}

In this section, we derive the pointing error distribution function and then use it to obtain the equivalent channel distribution function in the presence of pointing error. First, let's rewrite \eqref{he1} as:
\begin{align}
	\label{he4}
	h           =   h_L h_a h_p,
\end{align}
where $h_p= G_0(N) h'_p$ is the pointing error coefficient, and $0<h'_p<1$ is the normalized pointing error
\vspace{-.2 cm}
\begin{align}
	\label{he5}
	h'_p          = h'_{pt} h'_{pr} =
	\sqrt{G'_t(\theta_{tx},\theta_{ty})}
	\sqrt{G'_r(\theta_{rx},\theta_{ry})}.
\end{align}
From \eqref{he5}, $h'_p$ is a function of four RVs $\theta_{tx},\theta_{ty}$, $\theta_{rx}$, and $\theta_{ry}$.
The main lobe of the antenna pattern can be approximated by the Gaussian distribution function as:
\vspace{-.2 cm}
\begin{align}
	\label{b1}
		G_q(N)          = G_0(N) \exp\left(\! -\frac{\left( \tan^{-1}\left(\sqrt{\tan^2(\theta_{qx})+\tan^2(\theta_{qy})}\right)\! \right)^2}
	{w^2_B(N)}\right), 
\end{align}
where $G_0$ is defined in \eqref{cv}, and $w_B$ is the angular beamwidth (called the beam divergence) and which can be defined in few different ways.
It is observed that the calculation of $w_B$ based on $1/e$ criterion gives a better approximation, and can be obtained as \cite{balanis2016antenna}
\vspace{-.3 cm}
\begin{align}
	\label{s1}
	~~~G'_q(w_B, \phi_q)  - e^{-1}=0.
\end{align}
After performing a comprehensive search, for the standard array antenna introduced in \eqref{f_1}, $w_B$ can be approximated with a good accuracy as
$w_B(N) = \frac{B}{N}$ where $B=1.061$. The validity of this approximation is verified in Fig. \ref{m1}.
Moreover, based on the results of \cite{balanis2016antenna}, we have $G_0(N)<\pi N^2$ for array antennas.  As shown in Fig. \ref{m2}, for the standard array antenna, we can well approximate $G_0(N)\simeq \pi N^2$. In Fig. \ref{m1}, the validity of the Gaussian main-lobe approximation for standard array antenna is shown for different values of $\phi$. It is observed that changing $\phi$ affects only the side-lobes.

{\bf Theorem 1.}
{\it  The distribution of the pointing error is derived as}
\vspace{-.2 cm}
\begin{align}
	\label{f2}
	f_{h_p}(h_p) \!=\! \frac{\beta^2}{G_0^{\beta}} \ln\left(G_0\right) \times h_p^{\beta-1}
	    -\frac{\beta^2}{G_0^{\beta}} \ln\left(h_p\right) \times h_p^{\beta-1},
\end{align}
{\it where $0<h_p<G_0^2$ and $\beta=\frac{w_{B}^2}{\sigma_\theta^2}$.
For the standard array antenna, the pointing error distribution as a function of $N$ can be expressed as}
\vspace{-.2 cm}
\begin{align}
	\label{e7}
	f_{h_p}(h_p) &= \frac{B_1^2}{N^4 (\pi N^2)^{\frac{B_1}{N^2}}}
	\ln\left(\pi N^2 \right) \times h_p^{\frac{B_1}{N^2}-1} \nonumber \\
	&~~~-\frac{B_1^2}{N^4 (\pi N^2)^{\frac{B_1}{N^2}}}
	\ln(h_p) \times h_p^{\frac{B_1}{N^2}-1},
\end{align}
where $B_1=\frac{B^2}{\sigma_\theta^2}$.
\begin{IEEEproof}
	Please refer to Appendix \ref{AppA}. 
\end{IEEEproof}
%

{\bf Theorem 2.}
	{\it  The CDF of the pointing error is derived as}
	\vspace{-.2 cm}
\begin{align}  
	\label{e3}
	F_{h_p}(h_p) &= \left( \frac{h_p}{G_0}\right)^{\beta} - \beta    \left( \frac{h_p}{G_0}\right)^{\beta}     \ln\left(\frac{h_p}{G_0} \right).
\end{align}
{\it For the standard array antenna, the CDF of the pointing error as a function of $N$ can be expressed as}
\vspace{-.2 cm}
\begin{align} 
	\label{e4} 
	F_{h_p}(h_p) &= \left( \frac{h_p}{G_0}\right)^{\frac{B_1}{N^2}}
	\left[ 1  - \frac{B_1}{N^2}      \ln\left(\frac{h_p}{G_0} \right)   \right].
\end{align}
\begin{IEEEproof}
	Please refer to Appendix \ref{AppA}. 
\end{IEEEproof}
As can be seen, the pointing error model obtained for mmWave and THz links is different from that obtained for wireless optical systems in \cite[Eq. (11)]{farid2007outage}. 
The main reason for this difference is related to the difference in the design of mmWave/THz receivers as compared to optical receivers. In mmWave/THz systems, both the transmitter and the receiver have almost Gaussian patterns, but in optical systems, only the transmitted laser signal has a Gaussian pattern, and the aperture of optical receiver only collects the received Gaussian signal. 
Eqs. \eqref{e4} and \eqref{e7} are used to obtain the PDF and CDF of the pointing error for standard array antennas as a function of the number of antenna elements, $N$, which can be used to optimally design different considered systems/scenarios. Those equations can be further simplified using the below approximations.

{\bf Lemma 1.}
{\it  The PDF and CDF of the pointing error can be approximated as:}
\vspace{-.2 cm}
\begin{align}
	\label{e5}
	f_{h_p}(h_p) \simeq \frac{a\beta^2}{G_0} \left( \frac{h_p}{G_0}\right)^{\beta-1-\frac{1}{a}} - \frac{a\beta^2}{G_0} \left( \frac{h_p}{G_0}\right)^{\beta-1},
\end{align}
\vspace{-.5 cm}
\begin{align}
	\label{e6}
	F_{h_p}(h_p) \simeq \frac{a\beta^2}{\beta-\frac{1}{a}}\left( \frac{h_p}{G_0}\right)^{\beta-\frac{1}{a}} -  \frac{a\beta^2}{\beta} \left( \frac{h_p}{G_0}\right)^{\beta}.
\end{align}
\begin{IEEEproof}
	Please refer to Appendix \ref{AppA}. 
\end{IEEEproof}

In Theorems 1 and 2 as well as in Lemma 1, the pointting error models are provided for a high-directional mmWave/THz communication link between two unstable nodes. However, pointing error is one of the random parameters affecting high directional links. The other two important parameters are channel loss and small-scale fading.
Although channel loss is an important parameter in high-frequency communications, it is a static parameter and does not change the end-to-end PDF. However, the small-scale fading is a random parameter and changes the end-to-end PDF of the channel. Depending on the type of application, frequency and weather conditions, the small-scale fading model also changes. 
%
%
Several models have been used to model small-scale fading such as Rayleigh, Nakagami-m, Weibull, and $\alpha-\mu$ distributions \cite{priebe2011channel,7228163,kim2016statistical,9809917}.
In the following, we use the $\alpha-\mu$ distribution to model small-scale fading. It is a general distribution and includes several well-known models such as Rayleigh, Nakagami-m, and Weibull as special cases.


{\bf Theorem 3.}
{\it  The equivalent PDF of the considered THz channel is derived as}
\begin{align}
	\label{hn4}
	&f_h(h) = C_5     \left(C_2 h^\alpha \right)^{C_4}  e^{-\frac{C_2 h^\alpha}{2}} \Big[  \left(C_2 h^\alpha \right) ^{ - \frac{1}{2\alpha a} }  \\
	&\times
	\mathbb{W}_{  \frac{1}{2\alpha a}-C_3,  \frac{1}{2}+\frac{1}{2\alpha a}-C_3}  \left(C_2 h^\alpha \right)  
	-  
	\mathbb{W}_{  -C_3,  \frac{1}{2}-C_3 }  \left(C_2 h^\alpha \right)
	\Big],  \nonumber
\end{align}
{\it where }
$C_1 = \frac{\alpha \mu^\mu}{\hat{h_a}^{\alpha \mu} \Gamma(\mu)}$, 
$C_2 =  \frac{\mu}{\hat{h_a}^\alpha}$, 
$C_3 =\frac{\frac{\beta}{\alpha} -\mu+1}{2}$,
$C_4   = C_3  + \mu-1-\frac{1}{\alpha}$, and   
$C_5 = \frac{a\beta^2 C_1  C_2^{ - \mu+\frac{1}{\alpha}} }{\alpha  }$. {\it Moreover, the CDF of $h$ is derived as}
\begin{align} 
	\label{re3} 
	& F_h(h)  = 1-  \sum_{k=0}^{\mu-1} \frac{B_1 h}{\alpha}   \frac{a\beta^2}{\Gamma(k+1)}   
	\left[  
	(B_1 h)^{B_2} 
	e^{-\left( \frac{B_1 h)^\alpha }{2}\right)}  \right. \\
	&\left. \mathbb{W}_{B_3,B_4}\left( B_1^\alpha h^\alpha\right) 
	- 
	(B_1 h)^{B_5}
	e^{-\left( \frac{(B_1 h)^\alpha}{2}\right)}  
	\mathbb{W}_{B_6,B_7}\left( B_1^\alpha h^\alpha \right)     \right], \nonumber
\end{align}
where
\begin{align}
	\left\{ \!\!\!\!\! \! \!
	\begin{array}{rl}
		&B_1 = \frac{\mu^{1/\alpha} }{G_0 \hat{h_a} h_L}, ~~~
		B_2 = \frac{\alpha(k-1)}{2}+\frac{\beta}{2}-\frac{1}{2a}-1,~~~~~~~~ \\
		&B_3 = \frac{k-1}{2}-\frac{\beta}{2\alpha}+\frac{1}{2a\alpha}~~~
		B_4 = B_3+\frac{1}{2}, \\
		&B_5 = \frac{\alpha(k-1)}{2}+\frac{\beta}{2}-1~~~~
		B_6 = \frac{k-1}{2}-\frac{\beta}{2\alpha}~~~
		B_7 = B_6+\frac{1}{2}.
	\end{array} \right. \nonumber
\end{align}
\begin{IEEEproof}
	Please refer to Appendix \ref{AppB}. 
\end{IEEEproof}
Using the results of Theorem 3, the outage probability (i.e., the probability that the instantaneous signal-to-noise ratio (SNR) $\gamma=\frac{P_t h^2}{N_0}$ falls below a threshold $\gamma_{th}$) is obtained as follows:
	\vspace{-.2 cm}
\begin{align}
	\mathbb{P}_\text{out}= \int_0^{\gamma_{th}} f_{\gamma}(\gamma) \text{d}\gamma = F_h\left( \sqrt{{N_0 \gamma_{th}}\big/{P_t}}\right). \nonumber
\end{align}
\vspace{-.3 cm}

\section{Simulation and Numerical Results}
\vspace{-.1 cm}
In this section, we examine the accuracy of the provided analytical expressions by carrying out Monte-Carlo simulations using the standard array antenna pattern introduced in \eqref{f_1} and \eqref{cv}. To do this, we first generate $5\times10^7$ independent RVs $\theta_{tx}$, $\theta_{ty}$, $\theta_{rx}$, and $\theta_{ry}$. Then, for each independent run, using \eqref{f_1}, \eqref{cv}, and \eqref{f_2}, we generate $5\times10^7$ independent RVs of $h_p$. Next, using the generated instantaneous coefficients $h_p$, the PDF and CDF of the pointing error are obtained.

\begin{figure}
	\centering
	\subfloat[] {\includegraphics[width=2.5 in,height=1.5 in]{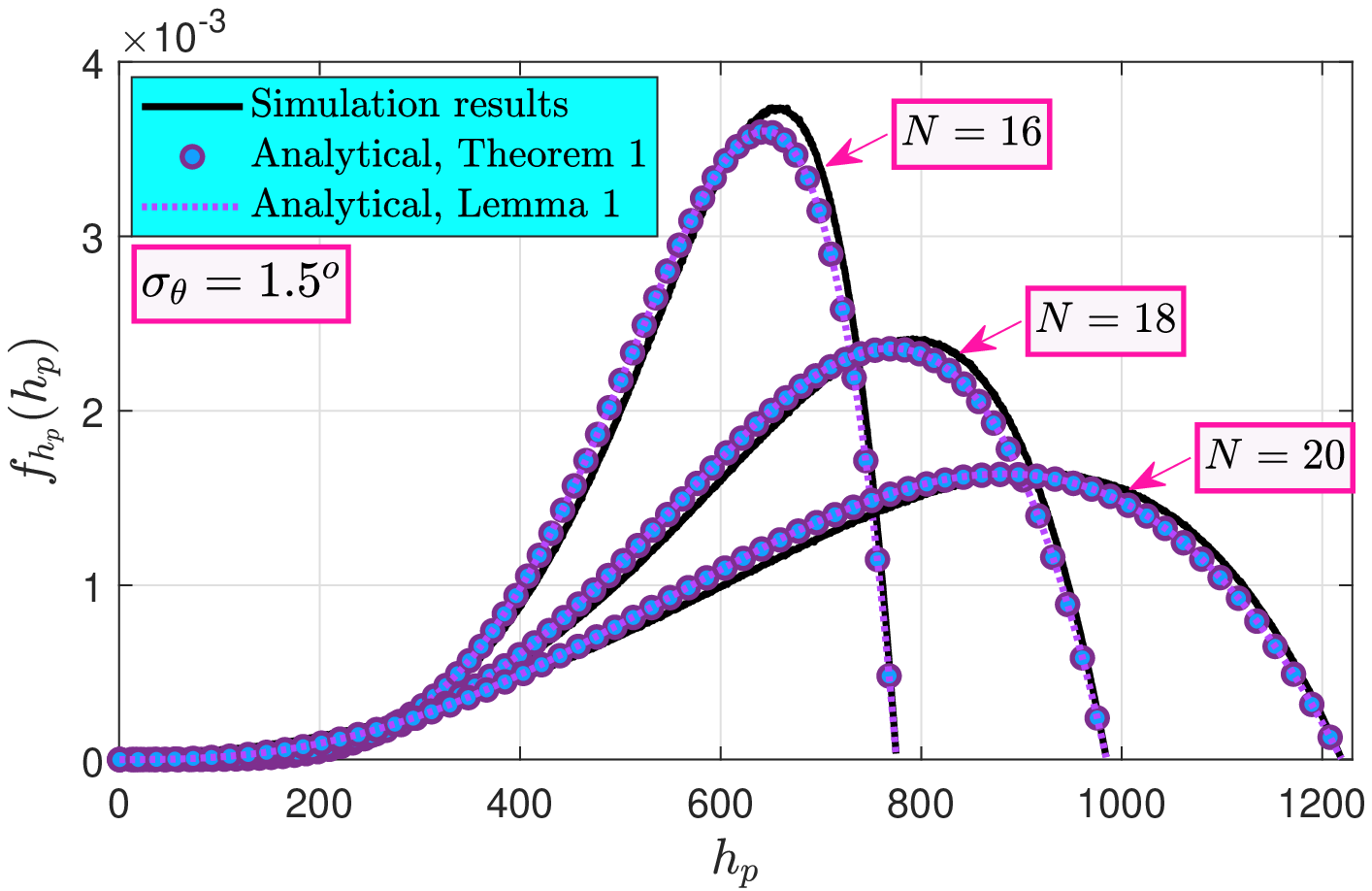}
		\label{gt1}
	}
	\hfill
	\subfloat[] {\includegraphics[width=2.5 in,height=1.5 in]{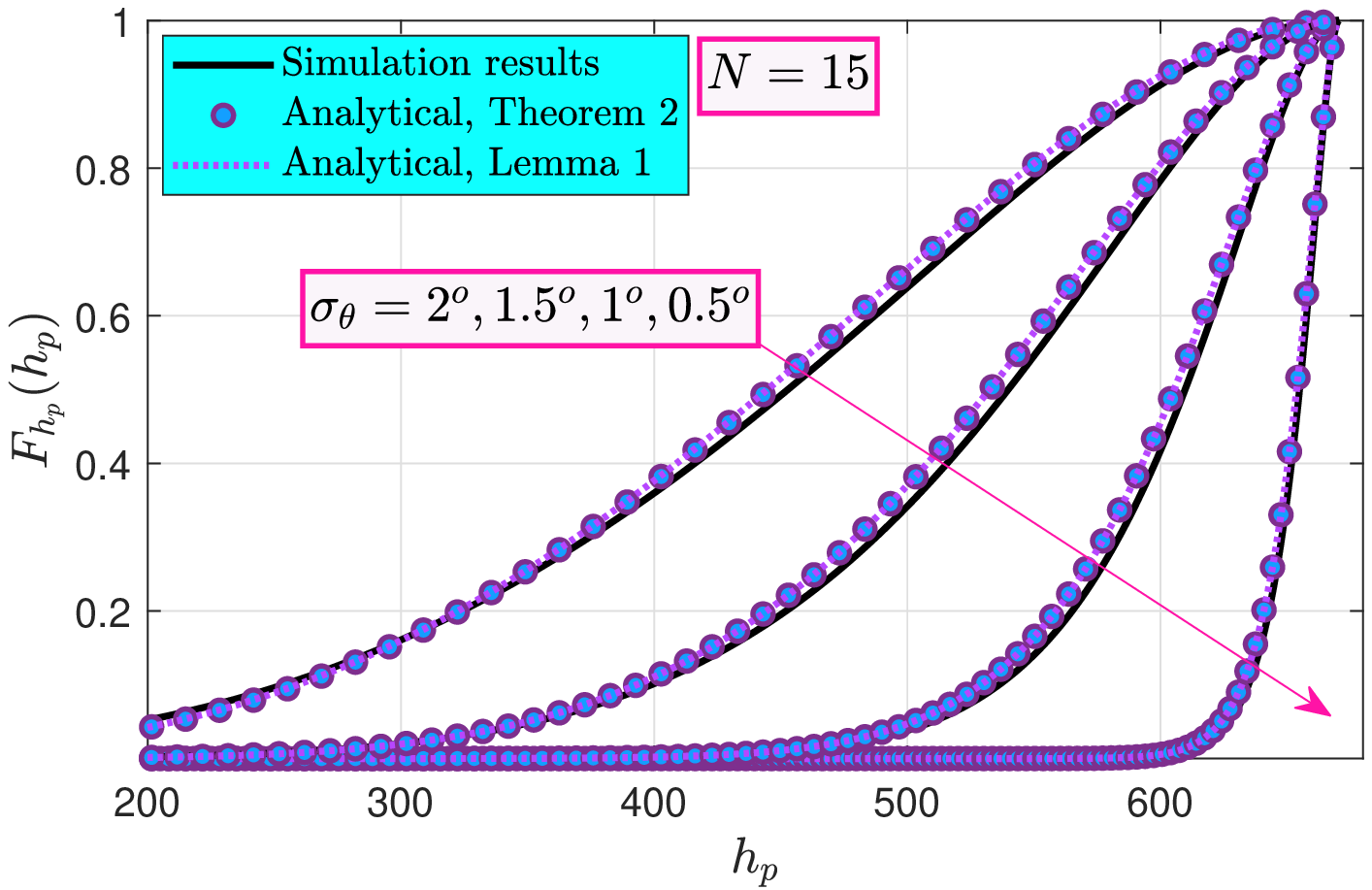}
		\label{gt2}
	}
	\caption{Comparison of the accuracy of the provided analytical  models for pointing error with the simulation results: (a) PDF of pointing error for different values of $N$; (b) CDF of pointing error for different values of $\sigma_\theta$.}
	\label{gt3}
\end{figure}

In Fig. \ref{gt1}, the PDF of the pointing error of a standard array antenna is plotted for different values of $N$.
The figure confirms the accuracy of the analytical expressions provided in \eqref{e7} and \eqref{e5}. 
The figure shows also that only by changing $N$ from 20 to 16, the channel distribution function changes significantly, and consequently, the system performance will change. 
This point clearly confirms the importance of choosing the optimal pattern in the presence of pointing error. While the antenna gain must be increased on one side to increase the SNR at the Rx, increasing the gain will make the system more sensitive to pointing errors. 
In Fig. \ref{gt2}, the CDF of pointing error is plotted for different values of $\sigma_\theta$. The results of this figure also confirms the accuracy of the analytical expressions provided in \eqref{e4} and \eqref{e6}.

\begin{figure}
	\begin{center}
		\includegraphics[width=2.5 in,height=1.5 in]{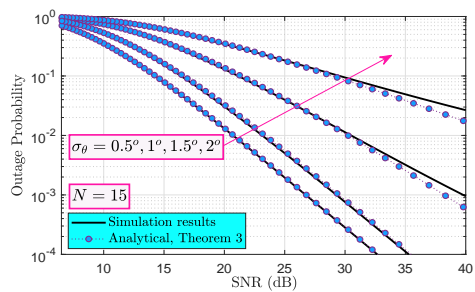}
		\caption{Comparison of the analytical results for the outage probability of the considered system at $f_c=275$ GHz with the simulation results for different values of $\sigma_\theta$.}
		\label{rn1}
	\end{center}
\end{figure}

Finally, in Fig. \ref{rn1} the end-to-end outage probability (related to the CDF of the equivalent channel) of the considered system at $f_c=275$~GHz and SNR threshold of 12~dB is depicted for four different values of $\sigma_\theta$. Again, simulation results confirm the  accuracy of the analytical results obtained from \eqref{hn4}.
As can be seen, for lower values of $\sigma_\theta$, the analytical results have a perfect match with the simulations. However, as  $\sigma_\theta$ is increased, a slight deviation between the analytical results and the simulations is observed. The reason for this is that with increasing $\sigma_\theta$, the effect of the side-lobes on the pointing error increases, which was not taken into account in the Gaussian pattern model.

%
%

\vspace{-.3 cm}
\section{Conclusion}
\vspace{-.2 cm}
As shown in this work, small angular errors affect high-directional mmWave/THz communications, and the larger the antenna gain, the more sensitive the system's performance is to the pointing error. Therefore, the main goal of this paper is to provide accurate models for the PDF and CDF of the pointing error. The validity of the analytical expressions was confirmed by simulations. The provided analytical expressions can be used for studying and designing high directional mmWave/THz systems.

\appendices

\vspace{-.2 cm}
\section{}
\label{AppA}
For small values of $x$, we have $\tan(x)\simeq x$, therefore, 
for lower values of $\theta_{qx}$ and $\theta_{qy}$, we can approximate RV $\theta_q$ as
\vspace{-.2 cm}
\begin{align}
	\label{he7}
	\theta_q = \sqrt{\theta_{qx}^2 + \theta_{qy}^2}.
\end{align}
Since $\theta_{qx}$ and $\theta_{qy}$ have Gaussian distributions, $\theta_q$ follows a Rayleigh distribution as
\vspace{-.2 cm}
\begin{align}
	\label{he8}
	f_{\theta_q}(\theta_q) = (\theta_q/\sigma_\theta^2)  e^{-{\theta_q^2}\big/{2 \sigma_\theta^2} }.
\end{align}
Using \eqref{he7}, Eq. \eqref{b1} can be approximated as
\vspace{-.2 cm}
\begin{align}
	\label{po0}
	G_q(N)          \simeq G_0(N) e^{ -{\theta_{q}^2} \big/{ w^2_B(N)}}.
\end{align}
Using \eqref{po0}, the instantaneous pointing error coefficient can be expressed as
\vspace{-.2 cm}
\begin{align}
	\label{po1}
	h_p          \simeq G_0(N) e^{ -{\theta_{tr}^2}\big/{2 w^2_B(N)}},
\end{align}
where $\theta_{tr} = \sqrt{\theta_{t}^2 + \theta_r^2}$. The distribution of RV $\theta_{tr}$ conditioned on $\theta_r$ can be obtained as \cite{papoulis2002probability}
\begin{align}  
	\label{er1}
	&f_{\theta_{tr}|\theta_r}(\theta_{tr}) = \frac{\text{d}}{\text{d}\theta_{tr}} \text{Prob}
	\left\{ \sqrt{\theta_{t}^2 + \theta_r^2}<\theta_{tr} \Big| \theta_r \right\} \nonumber \\
	&= f_{\theta_t}\left( \sqrt{\theta_{tr}^2-\theta_r^2} \right)    \frac{\text{d}\theta_t}{\text{d}\theta_{tr}}  
	= \frac{\theta_{tr}}{\sigma_\theta^2}  
	\exp\left(  -\frac{\theta_{tr}^2-\theta_r^2}{2 \sigma_\theta^2}   \right) .
\end{align}
Based on \eqref{er1} and \eqref{he8}, we have \cite{papoulis2002probability}
\begin{align} 
	\label{f1} 
	f_{\theta_{tr}}(\theta_{tr}) &= \int_0^{\theta_{tr}} f_{\theta_{tr}|\theta_r}(\theta_{tr}) f_{\theta_r}(\theta_r) \text{d}\theta_r
	 = \frac{\theta^3_{tr}}{2 \sigma_\theta^4}  
	e^{ -\frac{\theta_{tr}^2}{2 \sigma_\theta^2}  }.
\end{align}
Finally, using \eqref{he5}, \eqref{po1}, and \eqref{f1}, and after some manipulations, the distribution of the pointing error is derived  in \eqref{f2}.

Using \eqref{f1} and \cite[Eq. (2.33.13)]{jeffrey2007table}, the CDF of $\theta_{tr}$ is obtained as
\begin{align}
	\label{e2}
	&F_{\theta_{tr}}(\theta_{tr}) = 
	1 - 
	\exp\left(  -\frac{\theta_{tr}^2}{2 \sigma_\theta^2}   \right)         \left( \frac{\theta_{tr}^2}{2 \sigma_\theta^2} + 1  \right).
\end{align} 
From \eqref{po1} and \eqref{e2}, the CDF of the pointing error is derived as
\begin{align}  
	F_{h_p}(h_p) &= \text{Prob}\left\{ \theta_{tr}    >\sqrt{-2 w^2_B\ln\left(\frac{h_p}{G_0} \right)} \right\} \nonumber \\
	&= \left( \frac{h_p}{G_0}\right)^{\beta} - \beta    \left( \frac{h_p}{G_0}\right)^{\beta}     \ln\left(\frac{h_p}{G_0} \right).
\end{align}
Using \cite[Eq. (1.512.4)]{jeffrey2007table}, we can approximate the term $\ln(h_p)$ as $\ln(h_p) = \left(a h_p^{1/a}-a\right)$ where the parameter $a$ is a large number. In the simulation results, a value of $a=80$ is used.
Using this approximation in \eqref{f2} and \eqref{e3}, the PDF and CDF of the pointing error are derived in \eqref{e5} and \eqref{e6}, respectively.

\section{}
\label{AppB}
Using \eqref{he4} and \cite{papoulis2002probability}, the PDF of RV $h$ is obtained as
\begin{align}
	\label{hn1}
	f_h(h) = \int_0^{G_0}  \frac{1}{h_L h_p} f_{h_a}\left( \frac{h}{h_L h_p}\right)  f_{h_p}(h_p) \text{d}h_p.
\end{align}
Substituting \eqref{he3} and \eqref{e5} in \eqref{hn1}, applying a change of variables $y = h^\alpha h_p^{-\alpha}$, and after some manipulations, we have 
\begin{align}
	\label{hn2}
	&
		f_h(h) = \frac{a\beta^2 C_1}{\alpha C_2 } \left(\frac{h}{G_0 h_L}\right)^{\alpha \mu-\alpha-1}  
	\int_{C_2 \left(\frac{h}{G_0 h_L}\right)^\alpha}^\infty    
	\exp\left( -y\right)   \\
	%
	%
	&
	\times \left[    y^{-(C_3-\frac{1}{\alpha a})} 
	\left(C_2 \left(\frac{h}{G_0 h_L}\right)^\alpha \right)^{C_3-\frac{1}{\alpha a} }   
	- y^{-(C_3)}  
	\left(C_2 \left(\frac{h}{G_0 h_L}\right)^\alpha \right)^{C_3}
	\right] 
	dy,  \nonumber
\end{align}   
where
$C_1 = \frac{\alpha \mu^\mu}{\hat{h_a}^{\alpha \mu} \Gamma(\mu)}$, 
$C_2 =  \frac{\mu}{\hat{h_a}^\alpha}$, and
$C_3 =\frac{\frac{\beta}{\alpha} -\mu+1}{2}$.
In the following derivation, we use an integral identity \cite{jeffrey2007table}
\begin{align}
	\label{hn3}
	\int_u^\infty  x^{-\nu} {e^{-x}}  \text{d} x = u^{-\nu/2} e^{-u/2}  \mathbb{W}_{-\frac{\nu}{2},\frac{1-\nu}{2}}(u),
\end{align}
where $\mathbb{W}_{-\frac{\nu}{2},\frac{1-\nu}{2}}(u)$ is the Whittaker function. Now, using \eqref{hn2} and \eqref{hn3}, and after some manipulations, the PDF of $h$ is derived in \eqref{hn4}.

Using \eqref{he4} and \cite{papoulis2002probability}, the CDF of $h$ is obtained as
\begin{align}
	\label{re1}
	F_h(h) = \int_0^{G_0}    F_{h_a}\left( \frac{h}{h_L h_p} \right)  f_{h_p}(h_p) \text{d} h_p.
\end{align}
Based on \eqref{re1}, \eqref{e5}, \cite[Eq. (8)]{yacoub2007alpha}, and \cite[Eq. (8.352.2)]{jeffrey2007table}, and applying a change of variables $y = \mu\left( \frac{h }{\hat{h_a} h_L h_p} \right)^\alpha$, we obtain
\begin{align} 
	\label{re2} 
	& F_h(h)  = 1-  \sum_{k=0}^{\mu-1} \frac{B_1 h}{\alpha}   \frac{a\beta^2}{ \Gamma(k+1)}  \int_{(B_1 h)^\alpha}^\infty  
	\left[  (B_1 h)^{\beta-1-\frac{1}{a}} \right. \nonumber \\ &\left.
	\times y^{k-1-\frac{\beta}{\alpha}+\frac{1}{a\alpha}}e^{-y} - 
	(B_1 h)^{\beta-1} 
	y^{k-1-\frac{\beta}{\alpha}}e^{-y}   \right]  \text{d}y ,
\end{align}
where
$B_1 = \frac{\mu^{1/\alpha} }{G_0 \hat{h_a} h_L}$. 
Finally, using \eqref{hn3} and after some manipulations, the closed form expressions of \eqref{re2} is derived in \eqref{re3}.



\end{document}